\documentclass[10pt]{article}
\usepackage[letterpaper]{geometry}
\usepackage{hicss}
\usepackage[T1]{fontenc}
\usepackage[utf8]{inputenc}
\usepackage{times}
\usepackage{helvet}
\usepackage[protrusion=false,expansion=true]{microtype}
\usepackage{url}
\usepackage{latexsym}
\usepackage{booktabs}
\usepackage{array}
\usepackage{amsmath,amssymb}
\usepackage{indentfirst}
\usepackage{graphicx}
\graphicspath{{images/}}
\usepackage[round,authoryear,sort]{natbib}
\usepackage{float}
\usepackage{placeins}
\usepackage[hidelinks]{hyperref}
\usepackage{etoolbox}
\makeatletter
\patchcmd{\@maketitle}{1.31cm}{0.99cm}{}{\PackageError{hicss-final}{Title position patch failed}{}}
\patchcmd{\@makecaption}{\fontfamily{cmss}\selectfont}{\fontfamily{phv}\upshape\selectfont}{}{\PackageError{hicss-final}{Caption font patch failed}{}}
\g@addto@macro\UrlBreaks{\do\-}
\makeatother
\hypersetup{pdftitle={Auditing Institutional Heterogeneity for Generative AI in Patient Education: A Large-Scale Study of 102 US Transplant Handbooks},pdfauthor={Yubo Li; Rema Padman; Ramayya Krishnan},pdfkeywords={Generative AI, retrieval-augmented generation, patient education, clinical heterogeneity, LLM-as-judge}}
\makeatletter
\patchcmd{\@makecaption}{\fontfamily{cmss}\selectfont}{\fontfamily{phv}\upshape\selectfont}{}{\PackageError{hicss-final}{Caption body font patch failed}{}}
\makeatother

\makeatletter
\renewcommand{\paragraph}{\@startsection{paragraph}{4}{\z@}{-0pt plus -0.2ex minus -0.05ex}{-1em}{\normalfont\normalsize\bfseries}}
\makeatother

\title{Auditing Institutional Heterogeneity for Generative AI in Patient Education: A Large-Scale Study of 102 US Transplant Handbooks}

\author{Yubo Li \\
Carnegie Mellon University \\
\underline{yubol@andrew.cmu.edu}
\And
Rema Padman \\
Carnegie Mellon University \\
\underline{rpadman@andrew.cmu.edu}
\And
Ramayya Krishnan \\
Carnegie Mellon University \\
\underline{rk2x@andrew.cmu.edu}}

\date{}

\begin{document}
\maketitle

\begin{abstract}
Health systems are rapidly deploying generative-AI assistants that answer patient questions from institution-authored education materials, on the premise that grounding in local content yields consistent guidance. Do the underlying documents themselves agree? We use a structured-output large-language-model judge to audit 1{,}772{,}261 pairwise comparisons across 102 patient-education handbooks from 23 US solid-organ transplant centers, paired with 1{,}115 patient-derived questions (TransplantQA). Four findings bear directly on deployment: (1) same-center cross-organ agreement exceeds cross-center same-organ agreement by $0.024$ in the primary analysis (Holm-adjusted $p=0.011$), with sensitivity to document selection; (2) information gaps concern topics relevant to underrepresented subgroups, with reproductive health a \emph{double jeopardy}: 82\% absence and 86\% judge-rated high significance among divergent/contradictory pairs; (3) judge-derived themes form 991 clusters, with immunosuppression and pregnancy timing among the highest judge-rated priorities; (4) question and observed-coverage features predict high-divergence questions retrospectively (AUC $0.77$). We discuss implications for deploying patient-facing generative AI in transplant care.
\end{abstract}

\subsubsection*{Keywords:}

Generative AI, retrieval-augmented generation, patient education, clinical heterogeneity, LLM-as-judge

\section{Introduction}
\label{sec:intro}

Large language models show promise for medical question answering and patient education \citep{singhal2023medpalm}. Retrieval-augmented generation (RAG) can ground responses in institutional patient-education materials \citep{mace2025improving}. Grounding alone, however, does not establish whether the source documents cover a patient's question or provide consistent guidance. Measuring coverage and disagreement is therefore an important step in preparing these systems for deployment.

In transplant care, educational materials reflect different organs, care phases, patient populations, and local protocols. Literature describes variation in transplant access, eligibility policies, and immunosuppression management \citep{king2020major,nazzal2018center,lim2022heterogeneity}, alongside differences in patient-facing materials \citep{rivera2025examining,rodrigue2017readability,poudel2024readability,mace2025improving}. Such variation belongs to a longstanding area of clinical research \citep{wennberg1973small}. Differences in recommendations may be appropriate to their clinical contexts, but an AI assistant that combines sources without preserving those contexts can produce ambiguous guidance. An audit can identify these differences and questions that the available materials leave unanswered.

For health-IT teams, three operational questions follow. \emph{(i)~Auditing}: where are the coverage gaps and substantive differences in the corpus? \emph{(ii)~Triage}: which questions should receive priority for content or clinical review? \emph{(iii)~Mitigation}: what content updates, retrieval restrictions, and clinician checkpoints should the deployment include? These questions motivate a systematic measurement instrument for cross-institutional educational corpora.

This paper supplies that measurement. We audit 102 patient-education handbooks shared for this research by 23 U.S.\ transplant centers and obtained via a partnership with Transplants.org, a nonprofit supporting patients and families with information and education to navigate the transplant journey. This is paired with 1{,}115 patient-derived questions (TransplantQA), producing 1{,}772{,}261 answer-pair comparisons. A structured-output LLM judge assigns a relationship label, a free-text \emph{divergence topic}, and a \emph{clinical-significance} rating. These fields support review of individual differences alongside aggregate patterns; they do not establish clinical correctness.

We contribute: \textbf{(1)}~a large-scale audit of cross-institutional heterogeneity in U.S.\ transplant patient education; \textbf{(2)}~an exploratory comparison of institution and organ agreement patterns; \textbf{(3)}~coverage-gap analyses identifying potential information-equity concerns; \textbf{(4)}~a 991-cluster, judge-derived taxonomy of themes; and \textbf{(5)}~a retrospective predictive analysis of disagreement-prone questions (AUC $0.77$). The companion methodological paper \citep{li2026sourcedependence} describes HERO-QA and TransplantQA in full; here we focus on findings, judge calibration, and health-IT implications.

\section{Background and Related Work}
\label{sec:related}

\paragraph{Institutional heterogeneity in U.S.\ transplant care.}
Prior studies describe center-level variation in transplant access \citep{king2020major}, re-transplantation timing \citep{ku2025variations}, between- and within-center immunosuppression practices \citep{nazzal2018center,raghu2025association}, and eligibility policies \citep{lim2022heterogeneity,lee2023liver}. Research on patient-facing materials examines the availability of selection criteria \citep{rivera2025examining}, readability \citep{rodrigue2017readability,poudel2024readability}, and handbook comparison in a CMU--Transplants.org project report \citep{monahan2025handbooks}. We extend document comparison to a 1.77M-pair audit, adding structured judge metadata to identify recurring coverage gaps and disagreement themes.

\paragraph{Generative AI for patient education.}
Med-PaLM evaluated medical question answering and highlighted the need for further assessment before clinical use \citep{singhal2023medpalm}. The dominant benchmarks --- MedQA \citep{jin2021medqa}, MedMCQA \citep{pal2022medmcqa}, PubMedQA \citep{jin2019pubmedqa} --- assume a single gold answer and cannot surface inconsistency across the institutional sources a RAG system retrieves. RAG combines retrieved evidence with generation \citep{lewis2020retrieval}; retrieval quality and integration of evidence remain key challenges \citep{gao2023rag}, while reference-free quality scores such as RAGAS \citep{es2024ragas} leave the upstream question --- whether the corpus itself contains contradictions worth surfacing --- under-instrumented.

\paragraph{Clinical NLP for guideline comparison.}
Related clinical NLP work identifies disagreements between guidelines and new evidence \citep{borchert2021controversial}, retrieves evidence for guideline updates \citep{borchert2025highprecision}, builds annotated guideline corpora \citep{borchert2022ggponc}, and examines context behind apparent biomedical contradictions \citep{rosemblat2019characterization}. We extend this perspective to institutional patient-education materials, comparing the guidance an assistant retrieves across centers, organs, and care phases.

\paragraph{LLM-as-judge and healthcare-IT deployment audits.}
Since MT-Bench \citep{zheng2023mtbench}, LLM-as-judge has become standard practice, with dedicated judge models \citep{kim2024prometheus} and reference-free prompts \citep{liu2023geval}, most returning only a label or scalar score. Our structured-output variant adds divergence topics and clinical-significance ratings. Unlike RAGTruth \citep{niu2024ragtruth}, which evaluates hallucination against a reference, we compare answers grounded in different sources. Source faithfulness remains an assumption to check separately. Corpus audits can inform broader AI risk management \citep{nist_airmf_2023} and, where device regulation applies, change-control planning \citep{fda2025pccp}.

\section{Materials}
\label{sec:materials}

\paragraph{Corpus.}
The TransplantQA corpus contains 102 patient-education handbooks from 23 U.S.\ solid-organ transplant centers, covering 16 of the 20 largest U.S.\ programs by procedure volume. It spans heart (26), lung (26), kidney (22), liver (17), and pancreas (11) transplantation. For phase analyses, documents are grouped as pre-transplant (38), post-transplant (40), and combined (24), folding one evaluation document into pre-transplant and one numbered document into post-transplant. Public institutional PDFs supply the extracted text. Published exhibits use anonymous center and handbook codes, retaining organ and phase context.

\paragraph{Question set.}
The 1{,}115 questions were curated from public institutional Q\&A pages (31\%), community forums (25\%), patient-facing medical organizations (25\%), and other sources (19\%). Over 3{,}000 candidates were filtered for transplant relevance, duplication, and clarity; retained questions were lightly rephrased to remove identifying details and conversational dependencies. This is a curated public-source benchmark, not a prospective patient survey or a clinician-validated inventory of information needs. Topic frequencies reflect source selection rather than population demand.

The \emph{general} subset (311 questions) is posed to all 102 handbooks; the 804 \emph{organ-specific} questions are posed only to matching-organ handbooks. Each question has an organ label and one or more categories from a 13-topic taxonomy. Being posed to a handbook does not imply that its topic is within that document's intended clinical scope.

\paragraph{Generation and judging.}
HERO-QA \citep{li2026sourcedependence} uses hierarchical BM25 \citep{robertson2009probabilistic} and dense retrieval \citep{xiao2024c} with reranking, or the full extracted text for short handbooks ($\leq$80{,}000 characters). Qwen3-32B generates answers at temperature~0.1, returning \texttt{NOT ADDRESSED} when evidence is insufficient. Verification can trigger a full-document retry, subject to a context limit. A Qwen3-32B judge uses greedy decoding and structured JSON output.

\textsc{Absent} means at least one answer lacks relevant content; \textsc{Consistent} means the recommendations agree; \textsc{Complementary} means they add compatible details; \textsc{Divergent} means the recommendations differ substantively; and \textsc{Contradictory} means they directly oppose each other. The judge also supplies a divergence topic and a low/medium/high significance rating for disagreements. These ratings describe potential consequences inferred by the model, not observed patient harm.

\paragraph{Production run.}
The reference run contains 48{,}056 generated answers and 1{,}772{,}261 unique within-question handbook pairs. The comparison count is $311\binom{102}{2}+\sum_o N_q(o)\binom{N_h(o)}{2}$, where $N_q(o)$ and $N_h(o)$ are organ-specific question and handbook counts. Completed matrices record 1{,}583{,}176 pre-screened absent pairs and 189{,}085 judge calls; 156{,}876 pairs receive non-absent labels. Aggregating completed matrices avoids double-counting resumed execution. Generated-answer absence and answer-pair absence have distinct denominators.

\section{Analytical Methods}
\label{sec:methods}

Four analytical pipelines and a human-annotation study support the audit.

\paragraph{Theme clustering for the disagreement taxonomy.}
We extracted all 34{,}706 non-null \texttt{divergence\_topic} rows, deduplicated to 16{,}113 strings, embedded them with \texttt{all-MiniLM-L6-v2}, and applied average-linkage agglomerative clustering with cosine-distance threshold $0.35$. Each cluster is named by its most-frequent string; 237 of 991 contain $\geq\!10$ distinct variants. The extraction includes 3{,}561 \textsc{Complementary} and one \textsc{Consistent} record with non-null topics, so this is an exploratory map of judge-identified differences, not a clinically validated taxonomy of conflicting advice.

\paragraph{Agreement by institution and organ.}
Handbook pairs with $\geq\!20$ non-absent comparisons are grouped by shared institution and organ. Each pair's \emph{broad agreement} is the fraction labelled \textsc{Consistent} or \textsc{Complementary}; stratum means weight handbook pairs equally. Welch's two-sample $t$-tests allow unequal variances. We apply Holm correction across the six stratum contrasts and report unadjusted 95\% Welch confidence intervals for mean differences. Individual answer-pair labels are measurements, not millions of separate hypothesis tests. Because handbook pairs share documents and questions, these intervals and tests are exploratory and do not account for all dependence.

\paragraph{Per-handbook coverage z-scores.}
For each handbook $h$ and topic $t$, we compare absence to the pooled question--handbook baseline $\bar{r}_t$, using $z_{h,t}=(n_{h,t}^{\mathrm{abs}}-n_{h,t}\bar{r}_t)/\sqrt{n_{h,t}\bar{r}_t(1-\bar{r}_t)}$. Here $n_{h,t}$ counts applicable questions posed to the handbook, including unanswered ones. Values $|z|>2$ flag cells for descriptive review; they are not multiplicity-controlled significance claims. Topic overlap and shared questions limit a literal independent-binomial interpretation.

\paragraph{Predictive model.}
We define $y(q)=\mathbb{1}[r_{\mathrm{div}}(q)\geq Q_{75}]$, restricted to questions with $\geq\!30$ non-absent pairs ($n=441$). Features include organ and topic indicators, question length, 13 lexical flags, a 384-dimensional question embedding, and the observed number of non-absent pairs. We train gradient boosting and logistic regression on a stratified 75/25 question split (seed 42). AUC measures ranking of high-divergence questions; average precision summarizes precision across recall levels. The coverage feature makes this a retrospective audit model, requiring further validation for use on new queries.

\paragraph{Judge calibration.}
The completed human-annotation study contains 200 pairs, stratified as 40 per judge label, independently labelled by two annotators with judge outputs hidden and answers limited to 1{,}200 characters each. Inter-annotator agreement is 73.0\% ($\kappa=0.655$). Across all 200 pairs, judge agreement is 76.5\% with annotator A ($\kappa=0.706$) and 71.0\% with B ($\kappa=0.638$). On the 146 pairs where A and B agree, judge agreement is 87.7\% ($\kappa=0.842$; macro-F1 $0.841$). Per-label F1 on this subset is 1.00 for \textsc{Absent}, 0.83 \textsc{Consistent}, 0.70 \textsc{Complementary}, 0.69 \textsc{Divergent}, and 0.99 \textsc{Contradictory}. Excluding 54 disputed labels and oversampling rare contradictions make the subset score unsuitable as population accuracy.

Severity is less settled. Among 49 pairs with shared relationship labels and three available severity ratings, A--B agreement is 67.3\% ($\kappa=0.425$); judge agreement is 65.3\% with A ($\kappa=0.385$) and 93.9\% with B ($\kappa=0.827$). Table~\ref{tab:severity_validation} shows that medium-tier boundaries are particularly uncertain. We therefore use severity as a review flag, without treating either annotator as an adjudicated clinical reference.

\begin{table}[t]
\centering
\small
\caption{Judge severity matches by reference annotator's tier on the 49-pair subset. Denominators are reference-tier counts, not all sampled disagreements.}
\label{tab:severity_validation}
\begin{tabular}{lrr}
\toprule
Reference tier & Matches A & Matches B \\
\midrule
Low & 8/10 & 7/8 \\
Medium & 0/15 & 1/3 \\
High & 24/24 & 38/38 \\
\bottomrule
\end{tabular}
\end{table}

\section{Findings}
\label{sec:findings}

\subsection{Overall heterogeneity profile}
\label{sec:findings:profile}

Of 1{,}772{,}261 comparisons, 1{,}615{,}385 (91.1\%) are \textsc{Absent}. Among 156{,}876 non-absent pairs, 89{,}822 (57.3\%) are \textsc{Complementary}, 35{,}910 (22.9\%) \textsc{Consistent}, 31{,}057 (19.8\%) \textsc{Divergent}, and 87 (0.06\%) \textsc{Contradictory}. Explicit contradiction is rare; compatible differences in detail are more common. Separately, 35{,}365 of 48{,}056 generated answers (73.6\%) are absent. Pair absence is higher because either absent answer makes the pair absent.

Table~\ref{tab:per_organ} reports generated-answer absence $r_{\mathrm{abs}}$, the mean of question-level divergence rates $R_{\mathrm{div}}$ (among questions with at least one non-absent pair), and the share of questions with any divergent or contradictory pair. This question-weighted mean differs from the pooled pair percentages above.

\begin{table}[t]
\centering
\small
\caption{Per-organ heterogeneity from the production run. ``\% any div.'' is the share of questions in each organ with at least one divergent pair.}
\label{tab:per_organ}
\begin{tabular}{l rrr}
\toprule
 & $r_{\mathrm{abs}}$ & $R_{\mathrm{div}}$ & \% any div. \\
\midrule
general  & 0.778 & 0.175 & 55.6 \\
heart    & 0.692 & 0.146 & 45.3 \\
kidney   & 0.655 & 0.157 & 49.5 \\
liver    & 0.683 & 0.138 & 39.0 \\
lung     & 0.596 & 0.148 & 54.2 \\
pancreas & 0.668 & 0.185 & 29.9 \\
\bottomrule
\end{tabular}
\end{table}

Relative to an earlier Qwen3-14B hybrid-retrieval pipeline, the present system lowers generated-answer absence by $13.6$ percentage points averaged across organ groups, with mean $\Delta R_{\mathrm{div}}=-0.031$. Retrieval, generator, and judge all changed, so this system comparison cannot isolate retrieval effects or establish that all newly detected differences are source-authentic.

\subsection{Institution and organ agreement patterns}
\label{sec:findings:voice}

Do agreement patterns track organ boundaries, institutional affiliation, or both? We stratify 2{,}670 handbook pairs with $\geq\!20$ non-absent comparisons by shared center and organ (Table~\ref{tab:strata}). The primary analysis retains the production handbook identifiers; harmonizing three spelling variants of one center provides a sensitivity check below.

\begin{table}[t]
\centering
\small
\caption{Pair-stratum broad agreement. Same-center cross-organ pairs agree more than different-center same-organ pairs. ``same organ'' for same-center pairs means they differ only by care phase; the bottom row is the baseline.}
\label{tab:strata}
\setlength{\tabcolsep}{4pt}
\begin{tabular}{@{}l rr@{}}
\toprule
Stratum & Mean agree. & $n$ \\
\midrule
same center, same organ & 0.898 & 25 \\
\textbf{same center, different organ} & \textbf{0.861} & 100 \\
different center, same organ & 0.837 & 811 \\
different center, different organ & 0.782 & 1734 \\
\bottomrule
\end{tabular}
\end{table}

\begin{figure}[t]
\centering
\includegraphics[width=\linewidth]{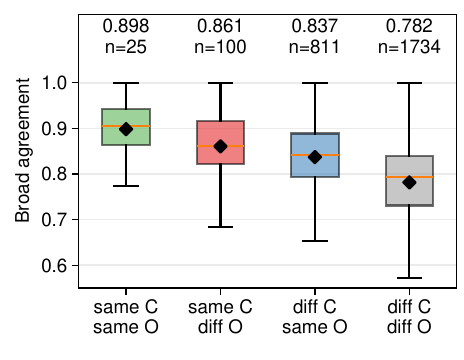}
\caption{Broad agreement by center (C) and organ (O). Same-center cross-organ pairs average 0.861 versus 0.837 for different-center same-organ pairs (Holm-adjusted $p=0.011$). Black diamonds mark means. Tests are exploratory because pairs share handbooks.}
\label{fig:stratum_box}
\end{figure}

The ordering in Table~\ref{tab:strata} and Figure~\ref{fig:stratum_box} suggests that both institutional affiliation and organ matching are relevant to agreement.

Same-center cross-organ agreement exceeds different-center same-organ agreement by $0.024$ (95\% Welch CI $[0.007,0.040]$; $t=2.82$; raw $p=0.0056$, Holm-adjusted $p=0.011$). For comparison, the same-organ versus different-organ contrast among different-center pairs is $0.055$. These contrasts do not decompose independent institution and organ effects, and the same-center same-organ stratum is small ($n=25$). The evidence supports a modest descriptive institutional association, not institutional dominance or a causal editorial effect.

Harmonizing center-name variants gives a similar contrast ($0.022$; 95\% CI $[0.005,0.039]$). However, the contrast is $-0.009$, $0.025$, and $0.050$ at minimum co-coverage thresholds of 10, 30, and 50, respectively. Restricting to answers generated from complete short-handbook text leaves 479 qualifying handbook pairs; the key contrast is $0.002$ (95\% CI $[-0.034,0.037]$; 23 same-center cross-organ versus 194 different-center same-organ pairs). Thus the institution contrast is sensitive to coverage and document selection. This subset removes passage-selection differences, but differs in corpus composition and cannot estimate a retrieval-error rate for the full run.

\subsection{Coverage gaps and patient information needs}
\label{sec:findings:coverage}

The \textsc{Absent} label identifies questions for which the pipeline produces no substantive answer from the supplied handbook text. We use these results to locate topic-level coverage gaps and identify priorities for improving the information available to an AI assistant.

\paragraph{Benchmark representation and coverage.} Reproductive Health is the second-largest topic in TransplantQA (291 questions, behind Medical Complications at 319), yet has the highest absence rate: 82.1\% of question--handbook answers are absent (Table~\ref{tab:topic_coverage}). Monitoring \& Follow-up has the lowest observed absence rate, at 63\%. These gaps indicate that the sampled handbooks alone provide limited answer coverage for the benchmark questions.

\begin{table}[t]
\centering
\small
\caption{Coverage and judge-rated significance across all 13 benchmark topics, ordered by absence.}
\label{tab:topic_coverage}
\setlength{\tabcolsep}{3pt}
\begin{tabular}{@{}>{\raggedright\arraybackslash}p{3.55cm} rrrr@{}}
\toprule
Topic & $N_q$ & Absent & High & $\leq$5\% cov. \\
 & & (\%) & (\%) & \\
\midrule
1 Reproductive health & 291 & 82.1 & 86.4 & 26/102 \\
2 Transplant process \& logistics & 21 & 78.6 & 95.0 & 41/102 \\
3 Special populations \& education & 73 & 78.5 & 84.4 & 17/102 \\
4 Financial \& insurance & 7 & 75.2 & 46.6 & 21/102 \\
5 Mental \& emotional health & 107 & 72.3 & 34.1 & 14/102 \\
6 Medical complications & 319 & 70.6 & 72.2 & 10/102 \\
7 Support \& community & 6 & 70.6 & 72.8 & 33/102 \\
8 Surgery \& recovery & 78 & 68.9 & 36.8 & 9/102 \\
9 Medications & 110 & 68.7 & 58.8 & 15/102 \\
10 Pre-transplant & 176 & 68.0 & 79.1 & 20/102 \\
11 Financial \& administrative & 19 & 65.9 & 70.7 & 28/102 \\
12 Lifestyle \& daily living & 225 & 64.7 & 55.8 & 11/102 \\
13 Monitoring \& follow-up & 108 & 63.1 & 73.6 & 6/102 \\
\bottomrule
\end{tabular}
\par\vspace{3pt}\raggedright
\emph{Note.} Absent uses question--handbook answers; High uses divergent/contradictory pairs with severity ratings. The last column counts handbooks with $\leq$5\% answer coverage on posed topic questions. Topic numbers identify points in Appendix Figure~\ref{fig:equity_supp}. Questions may have multiple topics; $N_q$ does not sum to 1{,}115. Across the 26 financial questions combined, 17/102 handbooks have $\leq$5\% coverage.
\end{table}

\paragraph{Patient needs requiring attention.} Coverage gaps concern reproductive-health, special-population, mental-health, and financial questions (Table~\ref{tab:topic_coverage}). These topics warrant targeted content review. Topic membership does not establish an asker's demographic identity or clinical eligibility; recipient age, organ and phase scope, and information provided through counseling or separate materials should inform which questions a handbook is expected to address.

\paragraph{Reproductive health as a content priority.} Alongside 82\% absence, 86\% of reproductive-health \textsc{Divergent}/\textsc{Contradictory} pairs are judge-rated high significance, compared with 34\% for mental health and 56\% for lifestyle (Appendix Figure~\ref{fig:equity_supp}, top). This is a severity estimate conditional on disagreement, not the probability that any two answers will disagree. The combination supports prioritizing pregnancy and medication guidance for coverage improvements and clinical review, subject to the severity-calibration limitations in Section~\ref{sec:methods}.

\paragraph{Care-phase differences.} Mean reproductive-health absence is 90\% in pre-transplant handbooks, 78\% post-transplant, and 76\% in combined documents (Appendix Figure~\ref{fig:equity_supp}, bottom). Per-handbook rates range from 53.4\% to 99.5\%, with 26 of 102 handbooks $\geq$95\% absent. Pre-transplant materials are therefore a useful starting point for reviewing whether reproductive-health information or links to appropriate counseling resources should be added.

\paragraph{Per-handbook spread.} Overall handbook absence ranges from 0.45 to 0.99, reflecting substantial variation in available answer coverage. Document scope, extracted text, and pipeline behavior all contribute to these measurements.

\subsection{Taxonomy of disagreement themes}
\label{sec:findings:themes}

The structured \texttt{divergence\_topic} field identifies the content of differences. Clustering yields 991 themes, 237 with $\geq\!10$ distinct variants (Section~\ref{sec:methods}).

\begin{table*}[t]
\centering
\small
\caption{Top 10 judge-derived themes by pair-instance count. \emph{High} is the high-significance share among instances with non-null severity ratings. Clustering also includes complementary differences.}
\label{tab:themes}
\begin{tabular*}{\textwidth}{@{\extracolsep{\fill}}lrrrr@{}}
\toprule
Theme (judge-derived) & Pair instances & Questions & Topic variants & High (\%) \\
\midrule
Post-transplant pregnancy timing             & 3{,}447 & 90 & 1{,}112 & 91 \\
Post-transplant blood-test frequency         & 2{,}224 & 17 & 816 & 84 \\
Organ-specific rejection symptoms            & 1{,}400 & 10 & 357 & 77 \\
Timing of routine dental care post-transplant & 1{,}064 & 12 & 298 & 43 \\
Safety of gardening post-transplant          & 1{,}030 & 6 & 215 & 78 \\
Antibiotic prophylaxis for dental procedures &   852 & 11 & 415 & 72 \\
Mycophenolate--pregnancy transition timing   &   810 & 24 & 306 & 95 \\
Timeline for resuming swimming               &   619 & 11 & 237 & 51 \\
Follow-up frequency and test schedule        &   616 & 4 & 81 & 88 \\
List of safe vaccines post-transplant        &   603 & 13 & 207 & 85 \\
\bottomrule
\end{tabular*}
\par\vspace{3pt}\raggedright
\emph{Note.} Questions and Topic variants count distinct benchmark question IDs and distinct judge-generated topic strings within each cluster, respectively.
\end{table*}

\begin{figure*}[t]
\begin{minipage}[t]{\columnwidth}
\vspace{0pt}
\centering
\includegraphics[width=\linewidth]{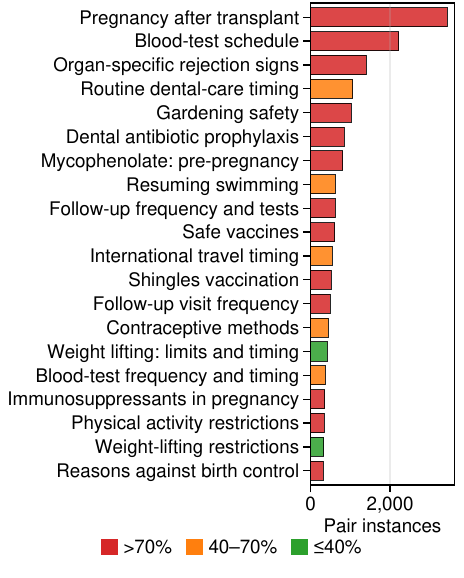}
\caption{Top 20 judge-derived themes by pair-instance count, coloured by the high-significance share among instances with severity ratings. These priorities reflect judge assessments, with the calibration limits reported in Section~\ref{sec:methods}.}
\label{fig:top_themes}
\end{minipage}\hfill
\begin{minipage}[t]{\columnwidth}
\vspace{0pt}
\centering
\includegraphics[width=\linewidth]{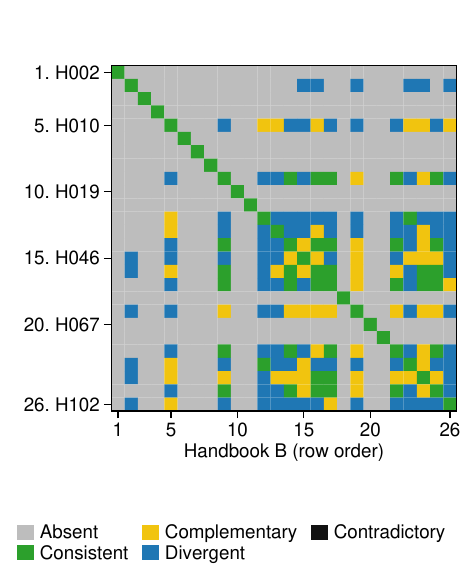}
\caption{Generated-answer comparison matrix for Q219 (dental care after transplant), across 26 heart handbooks. Colours encode judge labels: grey = Absent, green = Consistent, yellow = Complementary, blue = Divergent, black = Contradictory. Rows use anonymous handbook IDs; numbered columns follow the same order. Diagonal self-comparisons are excluded from counts; no off-diagonal cell is contradictory.}
\label{fig:heatmap_q219}
\end{minipage}
\end{figure*}

\paragraph{Judge-rated stakes.} High-significance shares vary across themes (Table~\ref{tab:themes}, Figure~\ref{fig:top_themes}): mycophenolate--pregnancy timing is rated high in $95\%$ of rated instances, pregnancy recommendations in $91\%$, and continuation of immunosuppressants during pregnancy in $98\%$ in a smaller cluster. Other shares include weight-lifting restrictions ($15\%$--$25\%$), gardening ($78\%$), and dental-care timing ($43\%$). These model-derived priorities require clinical adjudication before use in patient care.

\subsection{What predicts disagreement}
\label{sec:findings:predictive}

Among 441 questions with $\geq\!30$ non-absent pairs, 113 fall in the top quartile of divergence rate. Using the question and observed-coverage features in Section~\ref{sec:methods}, gradient boosting achieves test AUC $0.774$ and average precision $0.570$; logistic regression achieves $0.763$ and $0.541$. These results come from one question-level split of the audited corpus, without external-center or prospective validation.

Positive lexical coefficients include \texttt{qt\_can\_i} ($+0.85$) and \texttt{qt\_travel} ($+0.48$); negative topic coefficients include pre-transplant ($-0.58$), surgery/recovery ($-0.51$), and medications ($-0.43$). These are fitted associations, not intrinsic properties of question types. The model may help prioritize questions already covered by an audit; a query-time gate needs a model without observed-coverage features, calibrated thresholds, and prospective validation.

\subsection{Illustrative case studies}
\label{sec:findings:cases}

Table~\ref{tab:case_studies} illustrates judge-identified differences in follow-up schedules. Both examples compare different organs as well as institutions, so organ-specific monitoring needs are relevant to interpretation. These cases show why an assistant should retain the organ, care phase, and source institution when presenting guidance.

\begin{table*}[t]
\centering
\small
\caption{Illustrative differences in follow-up recommendations, labelled Divergent by the judge. Generated answers retain anonymous center codes, organ, and care phase for interpretation.}
\label{tab:case_studies}
\begin{tabular}{p{2cm} p{6cm} p{6cm}}
\toprule
Question & Handbook A & Handbook B \\
\midrule
Q1023: ``How often will I need follow-up appointments, and what kind of tests will be conducted?''
& \emph{C13 (heart, pre-transplant)} Weekly visits in the first year after a heart transplant, decreasing to every six months thereafter; cardiac focus including heart biopsies and echocardiograms.
& \emph{C02 (lung, post-transplant)} Weekly visits for first 4 weeks, bi-weekly for the next month, monthly through year~1, every 3 months thereafter; tests include UNOS-required ABGs, pulmonary function, and chest imaging. \\
\midrule
Q46: ``What is the recommended frequency for having blood tests done?''
& \emph{C12 (kidney, combined phases)} Weekly for the first month, then monthly thereafter; focus on kidney function and medication levels.
& \emph{C08 (lung, combined phases)} Mondays and Thursdays for the first 12 weeks, with additional comprehensive testing at specific intervals to adjust immunosuppressive medication. \\
\bottomrule
\end{tabular}
\end{table*}

\paragraph{Visualising one matrix in full.}
Figure~\ref{fig:heatmap_q219} shows Q219: ``How should I handle dental care following my transplant?'' Among the 325 heart-handbook pairs, 228 are absent, 17 consistent, 28 complementary, and 52 divergent. No pair is labelled contradictory in this production matrix. The example distinguishes missing information from substantive differences that warrant source-level review.

\section{Implications for Health IT Deployment}
\label{sec:implications}

The findings support four practical recommendations for transplant programs and health-IT teams preparing patient-facing AI systems.

\paragraph{Audit coverage and clarify differing guidance.}
First, inventory document versions, intended audiences, organs, and care phases. Use locally relevant questions to identify unanswered topics and differing recommendations. Clinicians should review original passages for a stratified sample of flagged pairs, including all severity tiers. Where differences persist after accounting for clinical context and pipeline errors, clarify applicable conditions or update guidance as appropriate. Assign responsibility for maintaining these clarifications and adjudicate severity before using the flags in deployment decisions.

\paragraph{Use the predictive layer for triage.}
Second, use predicted disagreement to order the retrospective review queue. For new-query routing, first remove audit-derived coverage features, validate on local queries, and choose thresholds based on clinician workload and the cost of missed disagreements. Candidate actions include clinician review, retrieval restricted to the treating program's applicable guidance, or disclosure that clarification is needed. AUC $0.77$ alone does not validate a top-decile routing rule or demonstrate improved patient outcomes.

\paragraph{Address priority content gaps.}
Third, prioritize reproductive-health, mental-health, special-population, and financial questions for content review. Transplant programs can improve coverage by adding concise guidance, linking relevant educational materials, and providing clear referral pathways for individualized questions. Pre-transplant reproductive-health information is a particular opportunity identified by this audit. Use local patient needs and clinician input to determine appropriate additions, assign an editorial owner, and recheck coverage after updates. Patient representatives can help assess whether the resulting resources address information-access barriers.

\paragraph{Retain evidence and monitor revisions.}
Fourth, retain source versions, retrieved passages, generated answers, judge outputs, and clinician decisions in an audit record. Model reasoning is an aid to review, not proof that a label is correct. Re-run affected questions after content or pipeline changes, monitor false alarms and missed disagreements, and test abstention and escalation behavior in a prospective pilot before broader deployment.

\section{Conclusion}
\label{sec:conclusion}

Across 1{,}772{,}261 comparisons of answers grounded in 102 handbooks from 23 centers, this audit identifies coverage gaps and differences relevant to patient-facing AI. Reproductive health is a priority for content improvement and clinical review; institution and organ patterns depend on which documents provide comparable evidence. Structured judge outputs and retrospective prediction can help target this work, while uneven severity agreement and retrieval sensitivity require validation. The findings support expanding coverage, clarifying differing recommendations, and monitoring the materials used to ground patient-facing AI.

\section*{Limitations}
\label{sec:limitations}

\paragraph{Scope and transferability.} Findings concern public, English-language U.S.\ transplant handbooks. The audit design could be adapted to oncology education, discharge instructions, or other multi-source patient materials, but would require representative local questions and a clinically calibrated rubric. International systems may differ in national standardization, local autonomy, language, and counseling pathways; neither effect sizes nor topic rankings should be transferred without replication.

\paragraph{Judge and annotation limitations.} Generation and judging use the same model family. The 200-pair study is small, label-balanced, and based on truncated answers; it does not validate every topic, institution, or severity tier. A saved same-model alternative protocol (label first, metadata afterward) has label $\kappa=0.669$ against production on these pairs, with 45 versus 80 divergent/contradictory labels. Because it also truncates answer inputs, this is evidence of protocol sensitivity, not a controlled prompt-only or cross-model validation. Independent-model judging and clinician adjudication remain necessary checks.

\paragraph{Snapshot and outcomes.} Handbooks were collected in 2024--2025. Guidance changes, and this corpus omits non-English materials and information delivered through other channels. No patient outcomes, demographic disparities, or deployed-system harms were measured; information-equity implications are hypotheses for follow-up assessment.

\paragraph{Retrieval and statistical dependence.} Apparent divergence can arise from extraction, retrieval, or generation errors. The complete-short-handbook analysis reduces passage-selection confounding only for that subset and does not reproduce the primary institution contrast. Controlled passage review is needed to attribute individual differences to source policy. Shared handbooks and questions also make the Welch tests exploratory despite Holm correction.

\paragraph{Institutional interpretation.} Center identities are withheld; anonymous codes preserve within-paper comparisons. Coverage and agreement scores describe the sampled handbooks and pipeline outputs; they are not measures of overall care quality. Document scope, patient populations, and information delivered through counseling or other materials should inform interpretation and any proposed content changes.

\section*{Acknowledgments}
We sincerely thank T. Mace and J. Mace of Transplants.org and the participating transplant centers for generously providing access to the patient-education handbooks, without which this research would not have been possible. This research was supported in part by the \href{https://ror.org/05xpvk416}{National Institute of Standards and Technology} under Federal Award ID 60NANB24D231 and by Carnegie Mellon University’s \href{https://www.cmu.edu/aimsec/research/index.html}{AI Measurement Science and Engineering Center (AIMSEC)}. This work used Bridges-2 at the Pittsburgh Supercomputing Center (PSC) through allocation CIS250181 from the \href{https://access-ci.org/}{Advanced Cyberinfrastructure Coordination Ecosystem: Services \& Support} (ACCESS) program, which is supported by U.S. National Science Foundation grants \#2138259, \#2138286, \#2138307, \#2137603, and \#2138296. We also gratefully acknowledge support from OpenAI’s Researcher Access Program, which provided API credits used in this research.

\FloatBarrier
\appendix
\begin{figure}[!ht]
\vspace*{1pt}
\section{Supplementary coverage analyses}
\label{app:equity}
\centering
\includegraphics[width=\linewidth]{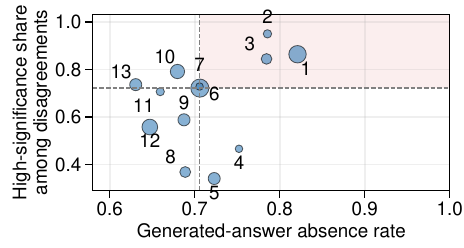}\\[4pt]
\includegraphics[width=\linewidth]{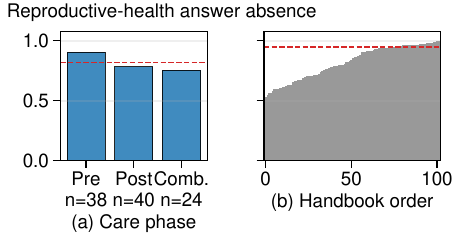}
\caption{Supplementary coverage analyses. Top: absence versus judge-rated high-significance share among divergent/contradictory pairs; topic numbers follow Table~\ref{tab:topic_coverage} and bubble size indicates question count. Shading denotes values above both medians, not a harm threshold. Bottom: reproductive-health absence by phase and handbook; 26/102 handbooks are $\geq$95\% absent (range 53.4\%--99.5\%).}
\label{fig:equity_supp}
\end{figure}
\FloatBarrier


\end{document}